\documentclass[aps,showpacs,amsmath,amssymbd,dvips,showkeys,preprintnumbers]{revtex4}

\usepackage{graphicx}   
\usepackage{color}      
\usepackage{braket}
\usepackage{tikz}
\usepackage{pgfplots}

\def \be  {\begin{equation}}
\def \ee  {\end{equation}}
\def \ee  {\end{equation}}
\def \bea {\begin{eqnarray}}
\def \eea {\end{eqnarray}}

\begin{document}
\preprint{ECTP-2019-07}    
\preprint{WLCAPP-2019-07}
\hspace{0.05cm}

\title{Polyakov linear-sigma model in mean-field approximation and optimized perturbation theory}

\author{Abdel Nasser Tawfik} 
\email{tawfik@itp.uni-frankfurt.de}
\affiliation{Nile University, Egyptian Center for Theoretical Physics (ECTP), Juhayna Square of 26th-July-Corridor, 12588 Giza, Egypt}
\affiliation{Goethe University, Institute for Theoretical Physics (ITP), Max-von-Laue-Str. 1, D-60438 Frankfurt am Main, Germany}

\author{Carsten Greiner} 
\affiliation{Goethe University, Institute for Theoretical Physics (ITP), Max-von-Laue-Str. 1, D-60438 Frankfurt am Main, Germany}

\author{Abdel Magied Diab}
\affiliation{Modern University for Technology and Information (MTI), 11571 Cairo, Egypt}
\affiliation{World Laboratory for Cosmology And Particle Physics (WLCAPP), 11571 Cairo, Egypt}

\author{M.T. Ghoneim}
\affiliation{Physics Department, Faculty of Science, Cairo University, 12613 Giza, Egypt}

\author{H. Anwer}
\affiliation{Physics Department, Faculty of Science, Cairo University, 12613 Giza, Egypt}

\date{\today}

\begin{abstract}

We compare results from the Polyakov linear-sigma model (PLSM) in optimized perturbation theory (OPT) with the mean-field approximation (MFA). At finite temperatures and chemical potentials, the chiral condensates and the decofinement order parameters, the thermodynamic pressure, the pseudo-critical temperatures, the subtracted condensates, the second- and high-order moments of various conserved charges (cumulants) obtained in MFA are compared with OPT and also confronted to available lattice QCD simulations. We conclude that when moving from lower- to higher-order moments of various quantum charges, OPT becomes more closer to QCD.       

\end{abstract}

\pacs{11.30.Rd, 11.10.Wx, 12.39.Fe, 02.60.-x}
\keywords{Chiral symmetries, Chiral transition, Chiral Lagrangian, Numerical approximation and analysis}

\maketitle

\tableofcontents
\makeatletter
\let\toc@pre\relax
\let\toc@post\relax
\makeatother 


\section{Introduction \label{intro}}

In theory of quantized fields \cite{Schwinger:1951xk,Schwinger:1953tb,Schwinger:1953zza,Schwinger:1953zz,Schwinger:1954zza,Schwinger:1954zz}, the linear-sigma model \cite{GellMann:1960np} with a spinless scalar field $\sigma_a$ \cite{Schwinger:1957em} and triplet pseudoscalar fields $\pi_a$ was introduced to describe the pion-nucleon interactions and the chiral degrees of freedom. This low-energy effective model has generators $T_a=\lambda_a/2$ with the Gell-Mann matrices $\lambda_a$ and a real classical field forming an O($4$) vector, $\vec{\Phi}=T_a(\vec{\sigma}_a, i\vec{\pi}_a)$. The chiral symmetry is explicitly broken by the $3\times 3$ matrix field $H=T_a  h_a$, where $h_a$ are the external fields. Under $SU(2)_L\times SU(2)_R$ chiral transformation, for instance, $\Phi\rightarrow L^+\Phi R$, $\sigma_a$ acquires finite vacuum expectation values, which in turn break $SU(2)_L\times SU(2)_R$ down to $SU(2)_{L+R}$. This results in massive sigma particle and light or nearly massless Goldstone bosons. The constituent quarks gain masses, as well, $m_q=g f_{\pi}$, where $g$ is the coupling and $f_{\pi}$ is the pion decay constant. Accordingly, fermions can be inserted in this model either as nucleons or as quarks. It has been shown that the $\sigma$ field under chiral transformations exhibits the same behaviour as that of the quark condensates, i.e $\sigma$ can be taken as an order parameter for the QCD chiral phase transition \cite{Birse:1994cz,Roder:2003uz,Gallas:2009qp,Tawfik:2014gga,Wesp:2017tze} and accordingly the phase structure \cite{Tawfik:2014gga,Tawfik:2016gye,AbdelAalDiab:2018hrx,Tawfik:2019rdd}, properties of QCD in finite magnetic fields \cite{Tawfik:2016lih,Tawfik:2016ihn,Tawfik:2017cdx,Tawfik:2019rdd} and various thermodynamic quantities can be estimated at finite bayon density \cite{Tawfik:2016gye,Tawfik:2016ihn,Tawfik:2016edq} and isospin asymmetry \cite{Tawfik:2019tkp}. 

Quantum mechanically, the spontaneous symmetry breaking could be achieved by  introducing a coherent state and minimizing the free energy density. Assuming that  the system of interest is enclosed in a cubic box and imposing periodic boundary conditions, the fields can be uniquely decomposed \cite{Randrup:1996es} into $\phi({\bf r},t)=\langle\phi(t)\rangle + \delta \phi({\bf r},t)=\sum_k\phi_k(t)\exp(i k \cdot {\bf r})$, where $\langle\phi(t)\rangle$ is the spacial expectation value over the box volume and $\delta \phi({\bf r},t)$ stands for the remaining fluctuations relative to a constant background field. It is obvious that the coefficient $\phi_k=\langle\phi \exp(-i k \cdot {\bf r})\rangle$ and the Fourier coefficients satisfy the symmetry relation, $\phi^{\ast}_k=\phi_{-k}$ because the fields, themselves, are real. Having all these, we can now apply the mean-field approximation (MFA), which is originated in statistical physics. In LSM, the meson fields are replaced by their spacial averaged values and all vacuum and thermal fluctuations are neglected. All quarks and antiquarks are retained as quantum fields. 

A generalization of MFA is the optimized perturbation theory (OPT), also known as $\delta$-expansion or variational perturbation theory \cite{Kunihiro:1983ej}. OPT was developed in $O(N)\; \phi^4$ theory and resums higher-order terms of the naive perturbation approach \cite{Thoma:1989ip,Thoma:1989in}. Here, we apply OPT on $O(4)\, \sigma$ model and then compare the results with the MFA. We aim determining the sensitivity of OPT relative to MFA.

The present script is organized as follows. A short review on the optimized perturbation theory and the mean-field approximation shall be given in section \ref{model}. Section \ref{resulat} is devoted to the results and discussion.  This includes chiral condensates and deconfinement order parameters, section \ref{Orders}, pseudo-critical temperatures,  \ref{Tmprs}, thermodynamic pressure, \ref{pressure}, fluctuations and correlations of conserved charges, \ref{fluctuations}. The latter are detailed to second-order, \ref{2ndfluctuations} and higher-order moments, \ref{Higherdfluctuations}. The final conclusions are outlined in section \ref{conclusion}.


\section{Optimized perturbation theory and mean-field approximation \label{model}}

We intend to check whether the optimized perturbation theory (OPT) would be able to play the role of an alternative to the nonperturbative approximation, such as the mean-field approximation (MFA), of the Polyakov linear-sigma model (PLSM). An OPT procedure aims at optimizing a linear $\delta$-expansion to the Lagrangian density. The basic idea of OPT becomes obvious when expanding the chiral Lagrangian, Eq. (\ref{deltaOPT}), from which we realize that even analytic nonperturbative calculations beyond what MFA would reach become accessible. The OPT procedure in PLSM goes as follows.
\bea
\mathcal{L}^\delta &=& (1-\delta)\;\mathcal{L}_0 (\eta) \;+\; \delta \;\mathcal{L} \;=\; \mathcal{L}_0 (\eta)  + \delta \; \Big[ \mathcal{L} - \mathcal{L}_0 (\eta)\Big], \label{deltaOPT}
\eea  
where $\eta$ is an arbitrary mass parameter, which can be fixed through an appropriate variational method \cite{Restrepo:2014fna}, and $\mathcal{L}_0 (\eta)$ is the free Lagrangian density in which $\eta$ is included. The parameter $\eta$ is equivalent to mass Thus, the implementation of OPT to PLSM is apparently accompanied by an expansion in terms of the arbitrary parameter $\delta$. Accordingly, Eq. (\ref{deltaOPT}) shows that the underlying symmetries in the chiral limit seems not modified. The term which is proportional to $\eta$ is added to the Lagrangian, while the same term multiplied by $\delta$ is also subtracted. At $\delta=1$, the original Lagrangian $\mathcal{L}$ can be recovered, straightforwardly. At $\delta=0$, the solvable Lagrangian $\mathcal{L}_0$ is obtained. In general, the parameter $\delta$ is also used as a dummy constant in order to label the orders of the perturbative calculations. Thus, $\delta$ is initially taken as a small value and afterwards fixed to unity. 

A short review on the LSM is now in order. LSM has chiral Lagrangian of $N_f$ quark flavors including the structure of mesons and quarks. With the incorporation of the Polyakov-loop potential, 
\bea
\mathcal{L} &=& \mathcal{L}_{\bar{\psi}\psi}+\mathcal{L}_m-\mathbf{\mathcal{U}}(\phi, \phi^{\ast}, T), \label{eq:plsmLgr}
\eea 
where the first term stands for the Lagrangian density for baryonic (fermionic) fields with $N_c$ color degrees-of-freedom, the second term gives the contributions of the mesonic (bosonic) fields, and finally the third term represents the Polyakov-loops potential incorporating the gluonic degrees-of-freedom and the dynamics of the quark-gluon interactions. 

When implementing OPT approach, Eq. (\ref{deltaOPT}), on the PLSM Lagrangian, Eq. (\ref{eq:plsmLgr}), we get
\bea
\mathcal{L}_{\bar{\psi}\psi} &=& \sum_f \overline{\psi}_f \Big[i\gamma^{\mu} D_{\mu}-\;\delta \;g\,T_a(\sigma_a + i \gamma_5 \pi_a) - (1-\delta)\,\eta\, \Big]\psi_f, \label{lfermion} \\
\mathcal{L}_m &=& \mathrm{Tr}\left[\partial_{\mu} \Phi^{\dag}\partial^{\mu} \Phi-\left(m^2 + \left( 1-\delta\right)\; \eta^2 \right) \Phi^{\dag} \Phi \right]  \nonumber \\ 
&+& \delta\, \left\{c\left(\mathrm{Det}\left[\Phi\right] + \mathrm{Det}\left[\Phi^{\dag}\right]\right)  - \lambda_1\, \left(\mathrm{Tr}\, \left[\Phi^{\dag} \Phi \right]\right)^2 
-\lambda_2\, \mathrm{Tr}\left[\Phi^{\dag} \Phi \right]^2
+\mathrm{Tr}\left[H\left(\Phi+\Phi^{\dag}\right)\right] \right\}.  \hspace*{6mm} \label{lmeson} \\
 \mathbf{\mathcal{U}}(\phi,\phi^{\ast}, T) & = & - b\, T \left[ 54\, \phi\, \phi^{\ast}\, e^{-a/T} + \ln\left(1-6 \phi \phi^{\ast} - 3\left(\phi \phi^{\ast}\right)^2+4 \left(\phi^3 + \phi^{{\ast}3}\right)\right)\right], \label{FUKU}
\eea
where $\psi$ are Dirac spinor fields for the quark flavors $f=[u,\, d,\, s]$, while $D_{\mu},\; \mu,\; \gamma^{\mu}$ and $g$ are covariant derivative, Lorentz index, chiral spinors, and Yukawa coupling constant, respectively. $\phi$ and $\phi^{\ast}$ are the order parameter of the Polyakov-loop variables and their conjugates, respectively. We note that that Polyakov Lagrangian, Eq. (\ref{FUKU}), isn't directly impacted by the OPT approach, for instance $\eta$ or $\delta$ isn't present, while $\mathcal{L}_{\bar{\psi}\psi}$ and $\mathcal{L}_m$ are impacted. We also notice that $\mathcal{L}_{\bar{\psi}\psi}$  is given in terms of $\eta$, while $\mathcal{L}_m$ of $\eta^2$. The reason can be understood due OPT approach and role of $\eta$, Eq. (\ref{deltaOPT}). It is obvious that $\eta$ as a modified mass parameter goes with the mass. Last but not least, when $\delta\rightarrow 0$, the {\it standard} PLSM Lagrangian \cite{Tawfik:2016ihn} can be obtained.

It should be noticed that the Lagrangian density of the fermions is deformed by adding a gaussian term $ \bar{\psi}(1-\delta)\eta \psi$ to the original Lagrangian density. All terms of coupling constant $g$ are multiplied by $\delta$. As discussed earlier, at $\delta \rightarrow 1$, the original Lagrangian density of the fermionic and mesonic contributions can be recovered. The generator operator $\Phi$ is a complex matrix for nonet meson states, $\Phi = \sum_{a=0}^{N_f^2 - 1} T_a \Big( \bar{\sigma_a} + \mathrm{i}\; \bar{\pi_a} \Big)$.  In U($3$) algebra, the generator operators $T_a = \hat{\lambda}_a/2$ are related to the Gell-Mann matrices $\hat{\lambda}_a$ \cite{Weinberg:1995mt}. The parameters $m^2$, $h_l$, $h_s$, $\lambda_1$, $\lambda_2$, and $c$ are determined, at sigma meson $m_\sigma=800~$MeV \cite{Schaefer:2008hk}. 

When using OPT instead of MFA to evaluate the free energy $\mathcal{F}$ of the PLSM, analytic nonperturbative calculations become possible through a prescription known as principle of minimal sensitivity (PMS) \cite{Stevenson:1981vj,Stevenson:1981rz}. PMS states that $\mathcal{F}$, Eq. (\ref{UPLSMtot}), can be minimized to the variations of $\eta$, at $\delta=1$
\bea
\left.\frac{\partial \mathcal{F}_{\mathtt{OPT}}}{\partial \eta}\right|_{\bar{\eta}, \delta=1}=0. \label{Gap-eq2}
\eea 
The expectation value of $\bar{\eta}$ is related to the sigma fields $\sigma_f$ and the color degrees of freedom $N_c$ as $\eta \sim \sigma_f$ \cite{Kneur:2010yv}. The grand canonical partition function $\mathcal{Z}$ which is given in dependence of the temperature $T$ and the chemical potentials of $f$-th quark flavor $\mu_f$ is defined by the path integral over all fermions, anti-fermions, and bosons 
\bea
\mathcal{Z}&=& \mathtt{Tr} \exp\left[\frac{\sum_{f=u,d,s} \mu_f \hat{\mathcal{N}}_f-\hat{\mathcal{H}}}{T}\right] 
= \int\prod_a \mathcal{D} \sigma_a \mathcal{D} \pi_a\int\mathcal{D}\psi \mathcal{D} \bar{\psi}\exp\left[ \int d^4 x
(\mathcal{L}+\sum_{f=u,d,s} \mu_f \bar{\psi}_f \gamma^0 \psi_f)\right],\hspace*{5mm}
\eea
where $\hat{\mathcal{H}}$ the chiral Hamiltonian density. The chemical potentials $\mu_f$ are related to the conserved charge numbers of baryon number ($B$), electric charge ($Q$) and strangeness ($S$) for each of the quark flavors, $f=[u,\;d,\;s]$
\bea
\mu_u = \frac{\mu_B}{3} + \frac{2 \mu_Q}{3}  , \qquad   \qquad
\mu_d = \frac{\mu_B}{3} - \frac{\mu_Q}{3} ,  \qquad   \qquad
\mu_s = \frac{\mu_B}{3} - \frac{\mu_Q}{3} -\mu_S. \label{muf-eqs}
\eea
Then, the free energy density can be deduced as
\bea
\mathcal{F}_{\mathtt{OPT}} (T,\;\mu_f) = - \frac{T}{V} \, \ln{[\mathcal{Z}]} =  \Omega(\sigma_l ,\; \sigma_s ) + \mathbf{\mathcal{U}}(\phi,\phi^{\ast}, T) + \Omega_{\bar{\psi}\psi}(T,\;\mu_f). \label{UPLSMtot}
\eea

\begin{center}
\begin{figure}[htb]
 \begin{tikzpicture}
\draw[black,thick,line width=2.0pt] (0,0) circle(1);    
    \node  at (0,-1.5) {(a)};
\end{tikzpicture}  
 \quad  \quad \quad
\begin{tikzpicture}
\draw[black,thick,line width=2.0pt] (0,0) circle(1);
      \draw[line width=2.0pt, black, dashed ]  (-1,0) -- (1,0);
\foreach \Point in {(-1,0), (1,0)}{
    \node at \Point {\textbullet};
}
    \node  at (0,-1.5) {(b)};
\end{tikzpicture} 
\quad  \quad  \quad
\begin{tikzpicture}
\draw[black,thick,line width=2.0pt] (0,0) circle(1);
   \draw[thick] (0,0) circle(0.25);
      \draw[line width=2.0pt, black, dashed ]  (-1,0) -- (-0.25,0);
      \draw[line width=2.0pt, black, dashed ]  (0.25,0) -- (1,0);
\foreach \Point in {(-1,0), (1,0), (0.25,0), (-0.25,0)}{
    \node at \Point {\textbullet};
}
    \node  at (0,-1.5) {(c)};
\end{tikzpicture}   
\caption{\footnotesize Diagrams illustrating the corrections to the free energy up to $\delta^2$-expansion. The fermionic contributions are depcited as solid lines. The propagators of sigma fields are represented by the dashed lines. \label{diagram} }
\end{figure}
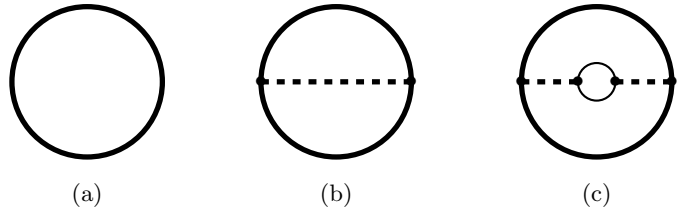
\end{center}

When recalling Feynman graphs up to $\delta^2$, the OPT approach can be illustrated. Fig. \ref{diagram} depicted the contributions in orders of $\delta$ and the color degrees of freedom $N_c$. The left panel (a) shows the zero order, $\delta^0$, with $1/N_c^0$ or $\mathcal{O}(\delta^0\; N_c^0)$, which is shown as thick solid lines representing the fermionic contributions, for which the system is composed of quarks and antiquarks. The free energy density of these contributions have been evaluated by different approximations including MFA \cite{ Schaefer:2009ui, Mao:2009aq, Tawfik:2016edq} and Hartree-Fock method (HFM) \cite{Klevansky:1992qe}. The middle panel (b) of Fig. \ref{diagram} shows the first-order corrections  of $\delta$ with $1/N_c^0$ or $\mathcal{O}(\delta\; N_c^0)$ and how this exceeds both MFA and HFM. The dashed line represents the propagator of the sigma field. The right panel (c) of Fig. \ref{diagram} shows the second-order corrections of $\delta$ with $1/N_c^0$ or $\mathcal{O}(\delta^2 N_c^0)$ belongig to the next-to-leading order (NLO) expansions \cite{Dmitrasinovic:1995cb, Oertel:1999fk}. 

We notice that the first two diagrams are the propagators up to $\mathcal{O}(\delta^2)$. These are considered to be an OPT approach for a gas with free fermions whose masses are converged by the modified mass parameter $\eta\rightarrow\eta+\delta (\sigma_f-\eta)$ \cite{Kneur:2010yv,Restrepo:2014fna}. The last diagram is already included in the second-order of $\mathcal{O}(\delta^2)$ and written with the usual mass parameter $\eta$ \cite{Kneur:2010yv,Restrepo:2014fna}. Let us now consider the first two diagrams to determine the PLSM free energy density in finite volume, $\mathcal{F}_{\mathtt{OPT}} (T,\;\mu_f)$, Eq. (\ref{UPLSMtot}), the {\it ''optimized''} $\delta$-expansion. 

As discussed, the full construction of the  OPT free energy density in finite volume is given in Eq. (\ref{UPLSMtot}). In rhs, the first term, $\Omega(\sigma_l ,\; \sigma_s ) $,  stands for the potential of the mesonic contributions in pure non-strange ($\sigma_l$) and pure strange ($\sigma_s$) condensates
\bea
\Omega(\sigma_l ,\; \sigma_s ) &=&    \Big(m^2\, + (1-\delta)\; \eta^2 \Big)\;\frac{ (\sigma^2_l+\sigma^2_s)}{2}  + \nonumber \\
&& \,\delta\; \Bigg\{ \frac{\lambda_1}{2} \, \sigma^2_l \sigma^2_s  +\frac{(2 \lambda_1
+\lambda_2)}{8}\, \sigma^4_l + \frac{(\lambda_1+\lambda_2)}{4} \, \sigma^4_s  - \frac{c}{2\sqrt{2}} \, \sigma^2_l \sigma_s - h_l \, \sigma_l - h_s\, \sigma_s \Bigg\}. \label{ULSMpot}
\eea  
We now show explicitly that the last term in rhs of Eq. (\ref{UPLSMtot}), $\Omega_{\bar{\psi}\psi}(T,\;\mu_f) $, which represents the potential of the quarks and antiquarks contributions and the chiral condensates in the mean-field limit reads  
\bea
\Omega_{\bar{\psi}\psi}(T, \mu _f)&=& -2 \,T \sum_{f=u, d, s} \int \frac{d^3\vec{P}}{(2 \pi)^3}  \Big[ \mathcal{I}_f^{(+)} (T, \; \mu_f) +\mathcal{I}_f^{(-)} (T, \; \mu_f) \Big]  \nonumber \\
&+& 2\,\delta\, N_c\, \sum_{f=u, d, s} \int     \frac{d^3\vec{P}}{(2 \pi)^3} \left\{\frac{(m_f+\eta)}{E_f}  (\eta -\sigma_f) \Big[ 1-n_f^{(+)} (T, \; \mu_f)-n_f^{(-)} (T, \; \mu_f) \Big] \right\}.  \label{PloykovPLSM}
\eea 
The Fermi-Dirac distribution functions $\mathcal{I}_f^{(+)} (T, \; \mu_f)$ and  $n_f^{(+)}(T, \; \mu_f)$ are defined - in a widely used notations - as
\bea
\mathcal{I}_f^{(+)} (T, \; \mu_f) &=& \ln\left[ 1+ 3\left(\phi\,+\phi^{\ast}\;e^{-\frac{E_f^{(+)}}{T}}\right) e^{-\frac{E_f^{(+)}}{T}}+e^{-3 \frac{E_f^{(+)}}{T}} \right] ,\\
n_f^{(+)}(T, \; \mu_f) &=& \frac{\left(\phi^{\ast} + 2\phi e^{-\frac{E_f^{(+)}}{T}}\right)e^{-\frac{E_f^{(+)}}{T}} + e^{-3\frac{E_f^{(+)}}{T}} }{ 1+ 3\left(\phi\,+\phi^{\ast}\;e^{-\frac{E_f^{(+)}}{T}}\right) e^{-\frac{E_f^{(+)}}{T}}+e^{-3 \frac{E_f^{(+)}}{T}}}, \label{Modfermi}
\eea
where $E_f^{(\pm)}=E_f\mp \mu_f$ are the energy-momentum dispersion relations, in which the upper sign is applied for quarks and the lower sign for antiquarks. These relations are subject of modifications due to OPT, $E_f= \left[|\vec{P}|^2 + (m_f +\eta)^2\right]^{1/2}$. By replacing $E_f^{(+)} $with $E_f^{(-)}$ and the order parameter of the Polyakov-loop variable $\phi$ with its conjugate $\phi^{\ast}$ or vice versa, we find that the terms $\mathcal{I}_f^{(-)} (T, \; \mu_f)$ and $n_f^{(-)}(T, \; \mu_f)$ are identical to $\mathcal{I}_f^{(+)} (T, \; \mu_f)$ and $n_f^{(+)}(T, \; \mu_f)$, respectively. It is worth highlighting that the second term in rhs of Eq. (\ref{PloykovPLSM}) is approximately equal the derivative of the first term with respect to mean-field of the averaged mass parameter $\bar{\eta}$. The second line in Eq. (\ref{PloykovPLSM}) is approximately the gap equation of the quark condensate \cite{Hansen:2006ee, Costa:2008dp}. By minimizing the thermodynamic potential with respect to the quark condensate $\langle \bar{q}_f q_f\rangle$, this definition becomes obvious, which shall play an essential role in clarifying the impacts of OPT compared to MFA, Fig. \ref{Fig.pressure}. The modified Fermi-Dirac distribution functions for quarks and antiquarks, Eq. (\ref{Modfermi}), with finite Polyakov-loop variables are derived explicitly by the summation over the Matsubara frequencies \cite{Hansen:2006ee}. These distribution functions straightforwardly lead to the standard form, especially in the limit that $\phi,\,\phi^{\ast}\rightarrow 1$, i.e. within the deconfined phase. On the contrary, within the confined phase, $\phi,\,\phi^{\ast}\rightarrow 0$, the exponential term grows by a factor of $3$. 

OPT and MFA procedures for determining the free energy density per unit volume in PLSM work as follows.
\begin{itemize}
\item {\it Optimized perturbation theory} (OPT): Eq. (\ref{UPLSMtot}) is to be estimated, where Eqs. (\ref{ULSMpot}), (\ref{PloykovPLSM}) and (\ref{FUKU}) shall be taken into account, and
\item {\it Mean-field approximation} (MFA): the chiral limit of the mean-field is obtained when the thermodynamic potentials in Eq. (\ref{UPLSMtot}) are determined at $\delta \rightarrow \;1$ and $\eta=0$ (vanishing arbitrary mass parameter). 
\end{itemize}
Having the free energy estimated weather in OPT or in MFA, the thermodynamic quantities which are thought to describe the chiral structure of the QCD matter at finite temperatures and finite chemical potentials are to be determined in OPT or in MFA. These are playing the role of the thermodynamic order parameters. Concretely, these are the mean $\sigma-$fields ($\bar{\sigma}_l,$ and $\bar{\sigma}_s$) and the Polyakov-loop variables ($\bar{\phi},$ and $\bar{\phi}^{\ast}$), which are estimated analytically from the global minimizing of the real part of the PLSM free energy density in finite volume, $\mathcal{R}e \;[\mathcal{F}_{\mathtt{OPT}} (T, \, \mu)]$, with respect to the associated thermodynamic order parameter, at a saddle point 
\bea
\left.\frac{\partial \mathcal{F}_{\mathtt{OPT}}}{\partial \bar{\sigma_l}}\right|_{\bar{\sigma_l}}=0, \quad 
\left.\frac{\partial \mathcal{F}_{\mathtt{OPT}}}{\partial \bar{\sigma_s}} \right|_{\bar{\sigma_s}}=0, \quad 
 \left.\frac{\partial \mathcal{F}_{\mathtt{OPT}}}{\partial \bar{\phi}} \right|_{\bar{\phi}}=0,\quad 
  \left.\frac{\partial \mathcal{F}_{\mathtt{OPT}}}{\partial \bar{\phi}^{\ast}}\right|_{\bar{\phi}^{\ast}}=0.  \label{Gap-eq}
\eea     
These expressions can be solved, numerically. In addition to Eqs. (\ref{Gap-eq}), we must assure   principle of minimal sensitivity (PMS), Eq. (\ref{Gap-eq2}), as well. Only when all expressions outlined in Eqs. (\ref{Gap-eq}) and (\ref{Gap-eq2}) are fulfilled, $T$ and $\mu$ can be determined and then enter further OPT calculations. The differences between OPT and MFA become now obvious, namely for MFA, $\mathcal{F}_{\mathtt{OPT}}\rightarrow \mathcal{F}$ and Eq. (\ref{Gap-eq}) are the only global minimization to be fulfilled. For OPT, Eq. (\ref{Gap-eq2}) must be fulfilled, as well.

In the section, that follows we present results on the chiral condensates and the deconfinement order parameters, section  \ref{Orders}. Section \ref{pressure} is devoted to the thermodynamic pressure. The fluctuations and correlations of conserved charges shall be discussed in section \ref{fluctuations}.  The second-order moments shall be outlined in section \ref{2ndfluctuations}, while the higher-order moments in section \ref{Higherdfluctuations}.

\section{Results and discussion \label{resulat}} 

In the present section, we study the quark-hadron phase structure of QCD matter at finite temperature and chemical potential  in MFA and OPT in the in PLSM. To this end, we first calculate the PLSM chiral condensates, $\sigma_l,$ and $\sigma_s$ and the deconfinement order parameters $\phi,$ and $\phi^{\ast}$. To judge about the reliability of both types of approximations, we compare the PLSM subtracted condensate for pure mesonic sigma fields with the lattice QCD (LQCD) calculations. In doing this, we can determine the pseudo-critical temperatures at different chemical potentials in MFA and then in OPT. The results in both approximations shall be summarized in the QCD phase diagram compared with recent LQCD simulations. As a further check, we discuss some PLSM features. For thermodynamics, we limit the discussion on the pressure. Then, we move to the fluctuations and the correlations of various conserved charges, which are deduced from the second derivatives of the pressure, for instance, with respect to the corresponding chemical potential. Last but not least, we compare between the PLSM results on high-order moments of particle numbers.

\subsection{Chiral condensates and deconfinement order parameters \label{Orders}}

As discussed in section \ref{model}, the chiral condensates ($\sigma_l$ and $\sigma_s$) and the Polyakov-loop variables ($\phi$ and $\phi^{\ast}$) can be evaluated by solving the gap equations, Eqs. (\ref{Gap-eq}) and (\ref{Gap-eq2}), $\mathcal{R}e \;[\mathcal{F}_{\mathtt{OPT}} (T, \, \mu)]$ at the subtle point. The vacuum values of nonstrange and strange chiral condensates at vanishing chemical potential $\mu_f=0~$MeV are $\sigma_{l0}=92.5~$MeV and $\sigma_{s0}=94.2~$MeV, respectively. The various PLSM parameters are estimated at $m_\sigma=800~$MeV. As introduced, for MFA, the gap equations to be solved are the ones given in Eq. (\ref{Gap-eq}).

\begin{figure}[htb]
\centering{
\includegraphics[width=15 cm, height=7 cm,angle=0]{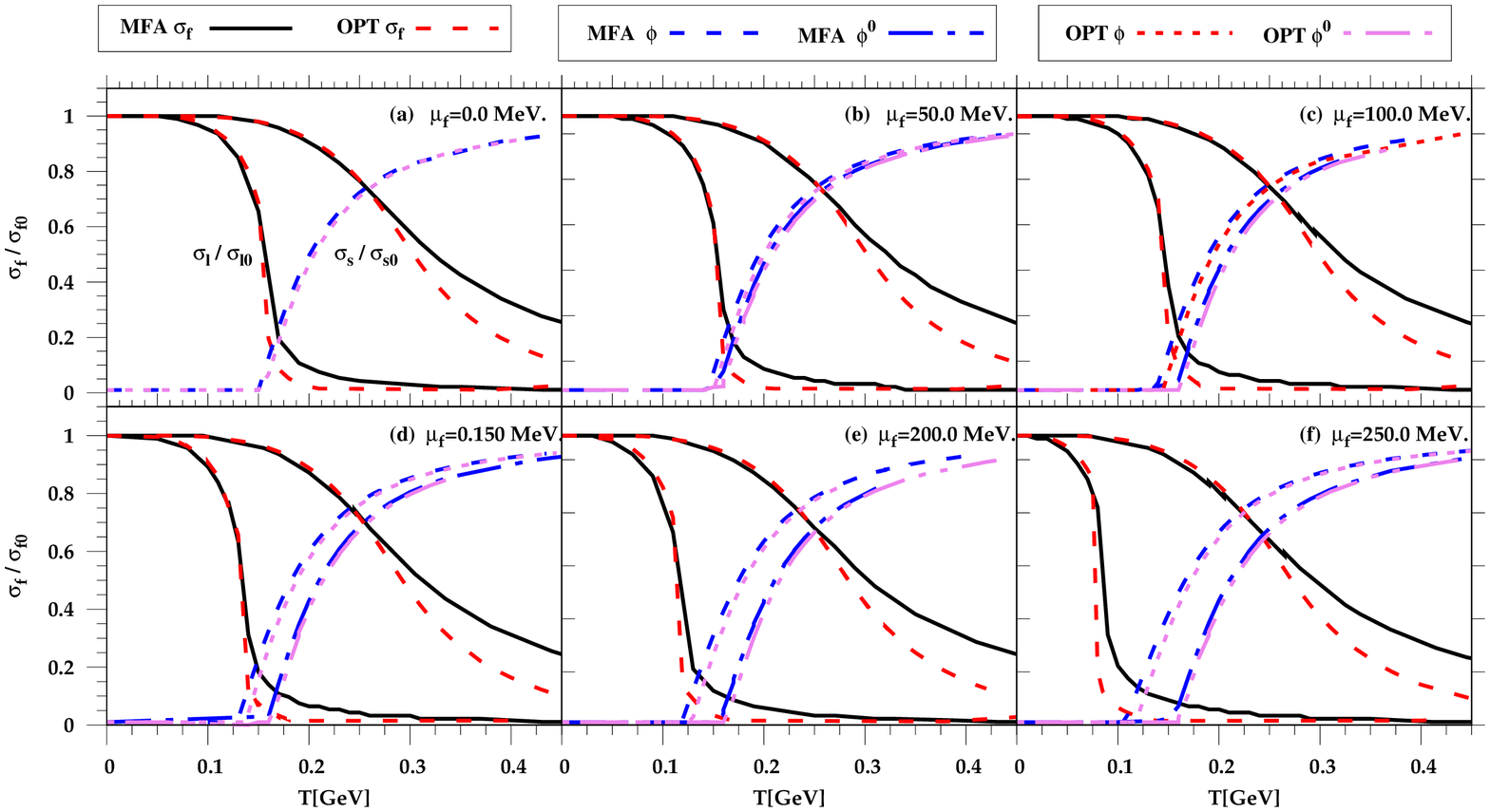}
\caption{\footnotesize (Color online) The temperature dependence of the normalized condensates for nonstrange $\sigma_l/\sigma_{l0}$ and strange $\sigma_s/\sigma_{s0}$ chiral condensates  in MFA (solid curves) and in OPT (dashed curves) and that of order parameter of Polyakov-loop variables  $\phi$ (dashed curves) and $\phi^{\ast}$ in MFA (dotted-dash curves) and  $\phi$ (dotted curves) and $\phi^{\ast}$ in OPT (double dotted dash curves) is given at $\mu_f=0$, $50$, $100$, $150$, $200$, and $250~$MeV in panel (a), (b), (c), (d), (e), and (f), respectively. 
\label{Fig.OrderPrameters}}
}
\end{figure}

Figure \ref{Fig.OrderPrameters} presents the temperature dependence of the order parameters which are calculated in MFA and OPT at different chemical potentials  $\mu_f=0$, $50$, $100$, $150$, $200$, and $250~$MeV in panels (a), (b), (c), (d), (e), and (f), respectively. The temperature dependence of the order parameter of the Polyakov-loop variables $\phi$ (dashed curves) and $\phi^{\ast}$ (dotted-dash curves) in MFA and  $\phi$ (dotted curves) and $\phi^{\ast}$ (double dotted dash curves) and in OPT are illustrated in the same panels at the given chemical potentials.

Between the MFA and OPT calculations for $\phi$ and $\phi^{\ast}$, there are small differences below and above $T_\chi$. While, there is almost no difference between the PLSM calculations for $\sigma_l$ and $\sigma_s$ in MFA and OPT, especially in the hadronic phase, at $T<T_{\chi}$, both approximations become distinguishable in the region of the phase transition and larger differences appear in the quark-gluon phase.  
\begin{itemize}
\item The thermal behavior of $\sigma_l$ and $\sigma_s$ starts at almost the same value, the one corresponding to their vacuum condensates. In this region, the effect of the chemical potential is apparently negligible. The increase in chemical potential seems to repair the increase in the phase transition. With this regards, we notice that OPT is more sensitive than MFA. 
\item We also notice that $\sigma_l$ in OPT starts its prompt phase transition more rapid than MFA. While $\sigma_l$ in MFA begins a {\it likely} first-order phase transition, at $\mu=200~$MeV, panel (e) of Fig. \ref{Fig.OrderPrameters}, $\sigma_l$ in OPT, indicates first-order phase transition, at $\mu=100~$MeV. 
\item  For $\sigma_s$, there is always a smooth phase transition. Although, OPT at $T\geq T_{\chi}$ is accompanied with a faster drop in $\sigma_s$ than MFA. The corresponding $T_{\chi}$ remains nearly identical, Fig. \ref{Fig.TmuDiagram}.
\end{itemize}

With this regard, it is in order now to summarize the procedures utilized in determining the pseudo-critical temperature $T_{\chi}$  \cite{Tawfik:2016edq}.
\begin{itemize}
\item The first one utilizes the temperature dependence of the chiral susceptibility. This is the second derivative with respect to the chemical potential or the first derivative of the subtracted condensate $\Delta_{ls}$, Eq. (\ref{subtracted}), with respect to the chemical potential, Eq. (\ref{fluct_correl}) and Fig. \ref{Fig.2nd_flac_corrl}. $T_{\chi}$ is precisely positioned at the inflection point, at which a maxima in the chiral susceptibility takes place.
\item The second one utilizes the intersection point of the normalized chiral condensate and the Polyakov-loop variables, as dictated by the LQCD simulations, Fig. \ref{Fig.OrderPrameters}, \cite{Cea:2003un}.  
\end{itemize}
With this regard, we believe that the first procedure leads to a precise estimation for $T_{\chi}$, at different chemical potentials. The pseudo-critical temperature $T_{\chi}$ is an essential thermodynamic quantity  for the QCD phase structure. In the section that follows, we confront our calculations on $\sigma_l$ and $\sigma_s$ to recent LQCD calculations.

\subsection{Pseudo-Critical Temperatures \label{Tmprs}}

\begin{figure}[htb]
\centering{
\includegraphics[width= 15 cm, height=5 cm, angle=0 ]{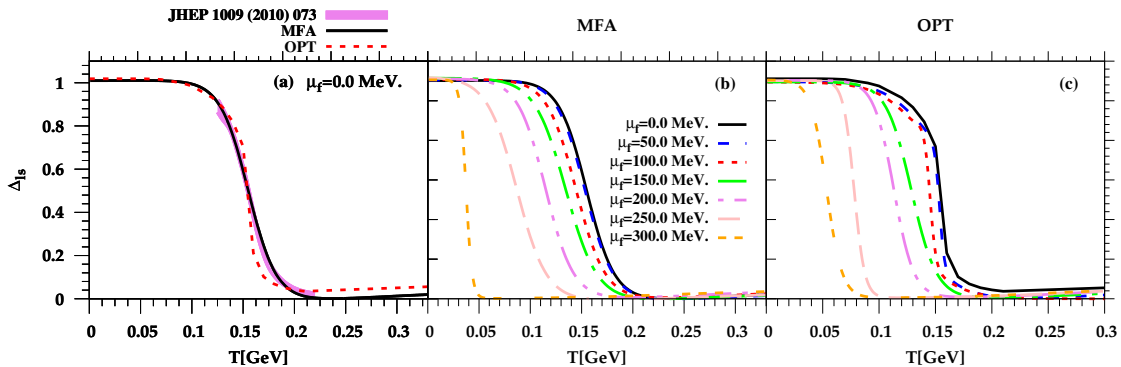}
\caption{\footnotesize (Color online) Left panel (a) illustrates the temperature dependence of the subtracted condensate $\Delta_{ls}$ in MFA (solid curve) and in OPT (dashed curve) and compares these with recent LQCD simulations \cite{Borsanyi:2010bp} (solid band). Middle panel (b) shows the MFA subtracted condensate at $\mu=0.0$ (solid curve), $50.0$ (dashed curve), $100.0$ (dotted curve), $150.0$ (dotted-dash curve), $200.0$ (double dotted-dash curve), $250.0$, (long-dash curve), and $300.0~$MeV. (dotted long-dash curve). Right panel (c) depicts the same as in the middle panel but here in OPT. 
\label{Fig.subsract}}
}
\end{figure}

Another thermodynamic quantity which plays the role of an order parameter is the normalized net-difference betwen the nonstrange and strange chiral condensate, known as the subtracted condensates $\Delta_{ls}$ \cite{Haas:2013qwp}
\bea
\Delta_{ls} =\frac{ \sigma_l -(h_l/h_s) \;\sigma_s \Big\lvert_{T}}{\sigma_l -(h_l/h_s) \;\sigma_s  \Big\lvert_{T=0}}, \label{subtracted}
\eea
where $h_l$ ($h_s$) are nonstrange (strange) explicit symmetry breaking parameters which are to be estimated from the Dashen-Gell-Mann-Oakes-Renner (DGMOR) relations \cite{GellMann:1968rz, Dashen:1969eg}. Fig. \ref{Fig.subsract} shows the temperature dependence of the subtracted condensate at different chemical potentials. The left panel (a) depicts the PLSM results in MFA (solid curve) and in OPT (dashed curve). These are compared with the continuum extrapolation of recent LQCD calculations \cite{Borsanyi:2010bp}. There is a good agreement, especially within the region of the phase transition, where the condensates decline very rapidly. It is obvious that at $T<T_\chi$, $\Delta_{ls}$ first remains unchanged having unity as a value. With the increase in temperature, a rapidly decrease takes place within the region of the phase transition. With further increase in the temperature, $\Delta_{ls}$ becomes almost temperature independent. In this region, the colored quark-gluon phase, $\Delta_{ls}$ keeps its low value almost unchanged, which apparently means that the quark and gluon are deconfined and their related degrees of freedom are liberated. 

The middle and right panels of Fig. \ref{Fig.subsract} show the temperature dependence of the subtracted condensates at various chemical potentials in MFA and OPT, respectively. Here, we also observe that at low temperatures, $\Delta_{ls}$ keep their values (unity) unchanged for a while, i.e. until the temperatures reach some values. Again, the further increase in the temperature gives almost the same behavior as the one obtained in the left panel (a). We notice that the increase in the chemical potentials tends to increase the rate of the rapid drop in $\Delta_{ls}$. The larger chemical potential, the narrower is the temperature region, within which $\Delta_{ls}$ decline to low value. It is apparent that hadron-quark phase transition of first order seems to start taking place, at $\mu_f = 200-250~$MeV. 

The results obtained when comparing $\Delta_{ls}$ with recent LQCD simulations, encourage mapping out the QCD phase diagram. This is another confrontation of the PLSM results in MFA and in OPT with the first-principle calculations. As discussed earlier, the pseudo-critical temperatures in PLSM, $T_{\chi}$ is to be estimated in different procedures. Having this done, we can now map out $T_{\chi}$ at different baryon chemical potential $\mu_B=3\,\mu_f$. 

\begin{figure}[htb]
\centering{
\includegraphics[width=7.5cm, height=12 cm, angle=-90]{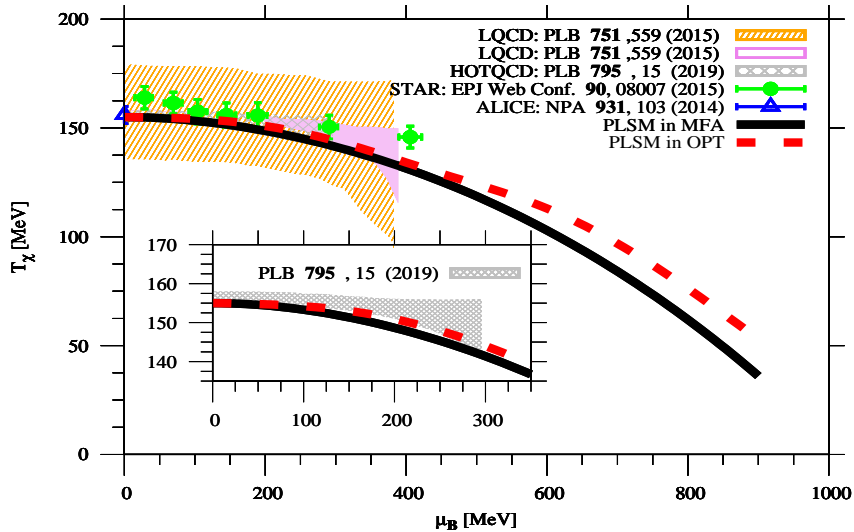}
\caption{(Color online)  The chiral phase-diagram of the PLSM resulted in MFA (solid curve) and in OPT (dashed curve) is compared with recent LQCD simulations \cite{Bellwied:2015rza}  (solid and dashed bands) and \cite{Bazavov:2018mes} (grid band). Furthermore, these are confronted to the experimental results of STAR \cite{Das:2014qca} (closed symbols) and ALICE experiment  \cite{Andronic:2017pug} (open symbols). 
\label{Fig.TmuDiagram}}
}
\end{figure}

Figure \ref{Fig.TmuDiagram} depicts the $T_{\chi}-\mu_B$ plane. The PLSM results in MFA (solid curve) and in OPT (dashed curve) are compared with the LQCD simulations \cite{Bellwied:2015rza} (dashed band), \cite{Bazavov:2018mes} (grid band) and also with the experimental estimations in the STAR experiment at RHIC (closed symbols) \cite{Das:2014qca} and the ALICE experiment at LHC (open symbols) \cite{Andronic:2017pug}. For the LQCD simulations \cite{Bellwied:2015rza}, the dashed band indicates the type of the phase transition. The grid band shows the same as the dashed band but here for the first-principle LQCD calculations \cite{Bazavov:2018mes}. While the solid band refers to the boundary of the critical temperature, the dashed band illustrates the boundary of $T_{\chi}$, which are obtained from the baryon susceptibility, Fig. \ref{Fig.2nd_flac_corrl}. 

The PLSM results in MFA (solid curve) and in OPT (dashed curve) agree well with both LQCD calculations. The agreement looks very convincing, especially when focusing on the LQCD simulations \cite{Bellwied:2015rza} (solid and dashed bands), which are recently refined \cite{Bazavov:2018mes} (grid band). The inside box zooms out the region, within which the LQCD calculations are reliable. When focusing on the possible differences between MFA and OPT, we would report on an almost identical result. Apart from the observation that OPT is accompanied with a slightly higher $T_{\chi}$, especially at large $\mu_B$, both approximations results in identical $T_{\chi}$. We notice that the phase transitions in MFA and in OPT are crossover. Both result in identical $T_{\chi}$, at $\mu_B\lesssim 450~$MeV. At larger $\mu_B$, they start being distinguishable; OPT has a higher $T_{\chi}$ than MFA.  

We conclude that at small $\mu_B$ both MFA and OPT are almost identical either in the order parameters or in the subtracted condensates or in the pseudo-critical temperatures. At higher $\mu_B$, OPT becomes more sensitive than MFA. 

In the section that follows, we compare PLSM results in OPT and in MFA with recent LQCD calculations for the thermodynamic pressure, which can be derived directly from the total free energy density, Eq. (\ref{UPLSMtot}).

\subsection{Thermodynamic Pressure} 
\label{pressure}

The thermodynamic pressure $p(T,\,\mu_f) = -\mathcal{F} (T, \mu_f)$ plays an important role in deriving various thermodynamic quantities. After addressing the PLSM in MFA and in OPT and estimating the corresponding order parameters [chiral (sigma fields) and deconfinement (Polyakov)], it is now comprehensible to analyze the thermodynamic pressure. As done in the previous section, we aim at confronting the PLSM results on the pressure with recent LQCD calculations \cite{Borsanyi:2016ksw}. The earlier have been estimated in MFA and in OPT, separately. We aim at determining which approximation agrees well with the LQCD calculations.

With this regard, we calculate the Stefan-Boltzmann (SB) limit which can be defined from the partition function of an ideal gas of free quarks and gluons \cite{Kapusta:2006book} with $N_c$ color degrees of freedom and $N_f$ quark degrees of freedom as
\bea
\frac{p_{\mathtt{SB}}}{T^4} &=& \frac{0.8\,G_g}{72} \;\pi^2 + \sum_f \frac{\,G_f}{72} \left[0.7 \;\pi^2 + 3 \left(\frac{\mu_f}{T}\right)^2+ \frac{3}{2\pi^2} \left(\frac{\mu_f}{T}\right)^4\right], \label{SBlimitEQ} 
\eea
where $G_g$ and $G_f$ are the degeneracy factors for gluons and quarks, respectively. These are defined as $G_g = \mathrm{spin \;polarization}[0,\,1] \times (N_c^2-1)$, $G_f= \mathrm{spin\; polarization}[+1/2,\,-1/2] \times  \mathrm{parity}[\psi ,\,\bar{\psi}] \times N_c N_f$, i.e. the degeneracy factors, $G_g=16$ and $G_f=36$. In this limit, i.e. taking into account the first two terms $\frac{\pi^2}{72}[(0.8 \times 16)+(0.7 \times 36)]$, the SB limit for thermodynamic pressure reads $p_{\mathtt{SB}}/T^4=5.209$ for $N_f=2+1$ quark flavors.

\begin{figure}[htb]
\centering{
\includegraphics[width= 15 cm, height=9 cm, angle=0  ]{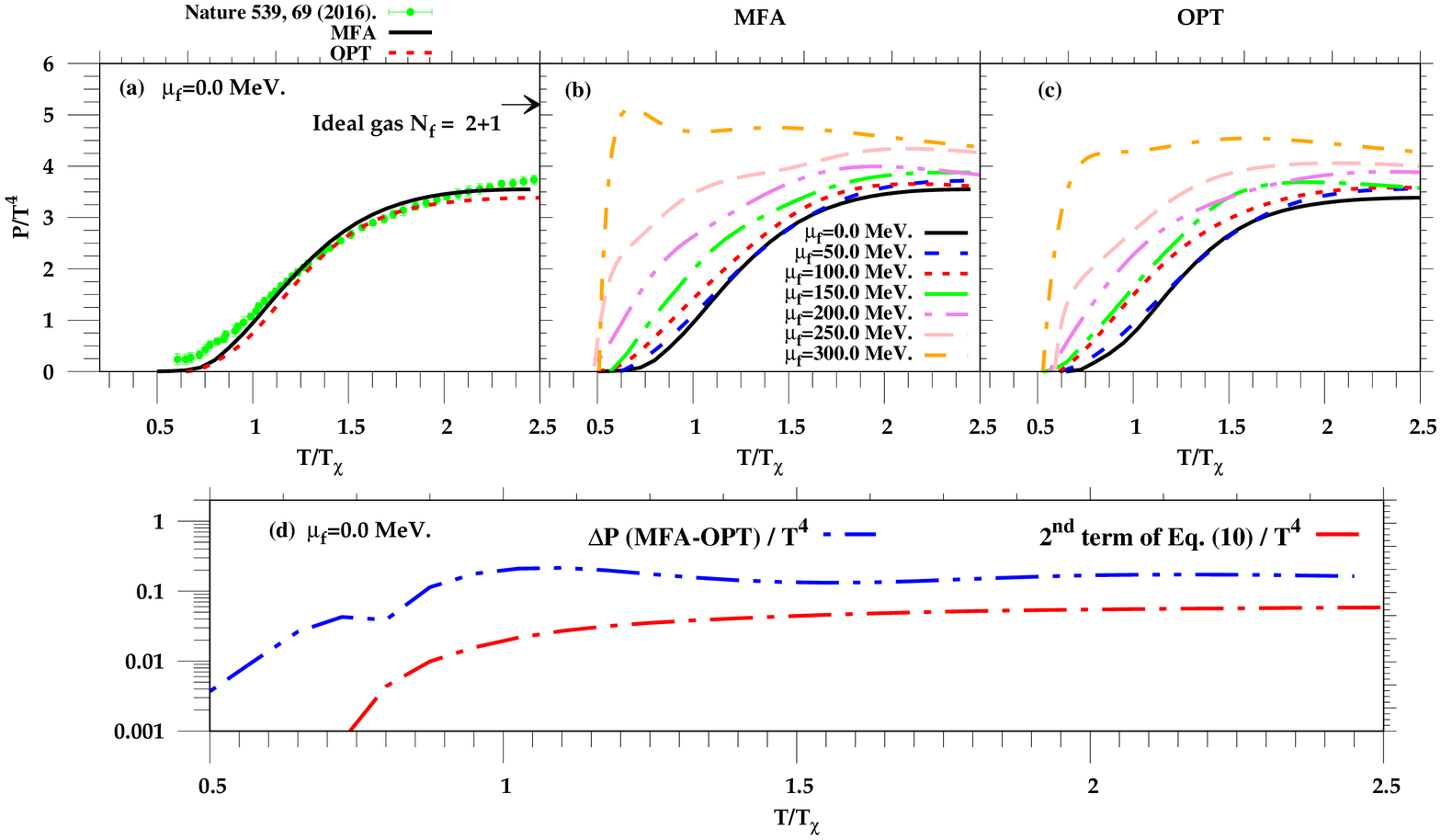}
\caption{\footnotesize (Color online) The same as in Fig. \ref{Fig.subsract} but here for normalized pressure $P/T^4$ (a-c). The PLSM results are compared with LQCD simulations \cite{Borsanyi:2016ksw}. The bottom panel (d) illustrates the temperature dependence of the PLSM in MFA and OPT on the normalized pressure (double-dot-dashed) and the second term in Eq. (\ref{PloykovPLSM}), the subtracted normalized pressure (dot-dashed curve).  
\label{Fig.pressure}} 
} 
\end{figure}

The upper panels of Fig. \ref{Fig.pressure} depict the temperature dependence of the normalised thermodynamic pressure $p/T^4$ deduced form the PLSM in MFA and OPT at different chemical potentials. The left-hand panel (a) shows the PLSM pressure in MFA and in OPT at $\mu_f=0~$MeV and compares these with recent LQCD calculations \cite{Borsanyi:2016ksw} (closed circles). We observe that the PLSM results agree well with the LQCD calculations. At low temperatures, the PLSM results are slightly lower than the LQCD calculations. The comparison is improved in the region of the quark-hadron phase transition. At  large temperatures, there is a good comparison, at least up to $T\leq 2.5 T_{\chi}$.  It is worthy emphasizing that the PLSM curves - similar to the LQCD - seem to saturate below the SB limit. At $T\leq 2.5 T_{\chi}$, the gap with the SB limit is about $31.8\%$ for MFA (solid curve) and a little bit more for OPT (dashed curve), $34.9\%$. In light of this, one  concludes that the phase transition in both approximations has the same transition (crossover), very similar to the LQCD calculations and the deconfined phase seems strongly correlated. 

Middle (b) and right (c) panels of Fig. \ref{Fig.pressure} show $p/T^4$ as functions of $T$ at $\mu_f=0$, $50$, $100$, $150$, $200$, $250$, and $300~$MeV in MFA and in OPT, respectively. We find that the increase in $\mu_f$ seems to drive the phase transition form being a smooth crossover to a prompt first-order phase transition.  In light of this, the pseudo-critical temperature $T_{\chi}(\mu_B)$ seems not a universal constant. At least, increasing $\mu_B$ decreases $T_{\chi}$, similar to the behavior depicted in Fig. \ref{Fig.TmuDiagram}. Not shown here, we report on other dependences, namely that $T_{\chi}$ depends - as well - on different variables, for instance, the type of the approximation used in the PLSM, the input parameters, and the Polyakov potential which is there to integrate the dynamics of gluon interaction to the PLSM chiral model.  

The bottom panel of Fig. \ref{Fig.pressure} (d) shows a semi-log scale of differences between OPT and MFA. The difference in the scaled pressure $\Delta p/T^4$ (double-dot-dashed) is compared with the difference in second term of Eq. (\ref{PloykovPLSM})  (dot-dashed curve). The latter represent main contributions added by the OPT approximation, i.e. extra term relative to MFA. We conclude that both approximations MFA and OPT seem not giving  the same thermodynamic pressure. This would not be obvious from a blind comparison as in the top panels. A great portion is to be credited to the second term of Eq. (\ref{PloykovPLSM}).

In the section that follows, we move to high-order moments. In doing this, we aim at highlighting whether the higher-order moments enhance the difference between MFA and OPT, as observed in Fig. \ref{Fig.pressure} or eventually not. We first present calculations quadratic fluctuations for the same quantum charges and correlations, i.e. mixed quantum charges, section \ref{2ndfluctuations}. The higher-order moments shall be elaborated in section \ref{Higherdfluctuations}.

\subsection{Fluctuations and Correlations of Conserved Charges}
\label{fluctuations}

The fluctuations and correlations of various conserved quantum charges, $X=[B,\, Q,\,S]$, can be estimated from the derivative of the total free energy density of the system of interest, Eq. (\ref{UPLSMtot}), with respect to the associated chemical potential, Eq. (\ref{muf-eqs}),
\bea
\chi^{BQS}_{ijk} &=& \frac{\partial^{i+j+k}\; (p(T, \hat{\mu}_X)/T^4)}{(\partial \hat{\mu}_B)^i\, (\partial \hat{\mu}_Q)^j\, (\partial \hat{\mu}_S)^k\,}, \label{fluct_correl}
\eea
where $ \hat{\mu}_X = \mu_X/T$, the superscripts $i$, $j$ and $k$ run over integers giving the orders of the derivatives. With this regard, it is informative to estimate the fluctuations (diagonal) and correlations (off-diagonal) of PLSM in MFA and in OPT. The thermal expectation values of the conserved charges $X=[B,\, Q,\,S]$, first-order moments, are estimated from the derivative of the partition function $\mathcal{Z}(T,\,\mu_X)$ with respect to corresponding chemical potential $\mu_X$ as 
\bea
\left\langle N_X \right\rangle &=&T\,\frac{\partial \ln\Big[\mathcal{Z}(T,\,\mu_X) \Big]}{\partial \mu_X}.
\eea
The cumulants of the quantum number distributions are given as 
\bea
C_n^X &=& V T^3 \chi_n^X, 
\eea
where  $\sigma^2 =\left\langle (\delta N^2) \right\rangle = VT^2 \chi_2^X $ is the variance and  $\kappa = C^X_4/(\sigma^2)^2$ is the kurtosis. One can construct the moment products, which can be related to the measured multiplicities of produced particles \cite{Luo:2017faz}, for instance, 
\bea
\kappa \sigma^2 &=& \frac{C_4^X}{C_2^X} = \frac{\chi_4^X}{\chi^X_2}, \label{kappsigma2}
\eea
i.e. ratios of quartic to quadratic cumulants of the net-quantum number fluctuations. The fluctuations of conserved  quantum charges can be determined in the SB limit, i.e. for an ideal gas with free constituents. These are listed in Tab. \ref{tab:2}, where only up to the $4$-th order cumulants are finite. Comparing these values with our calculations indicates how far is the deconfined system from the SB limit.

\begin{table}[htb]
\begin{center}
\begin{tabular}{| c | c | c | c |} 
\hline
 $\mathtt{Cumulants}\;$ $\mathtt{in}\;$ $\mathtt{SB}\;$ $\mathtt{limit}$ & $B$ & $Q$ & $S$  \\ \hline
\hline 
$\chi_2^{X}$ & $1/3$ & $2/3$ & $1$\\ 
\hline 
$\chi_4^{X}$ & $2/9 \pi^2$ & $4/3 \pi^2$ & $6/ \pi^2$\\ 
\hline 
\end{tabular}
\caption{Cumulants for baryon ($B$), electric charge ($Q$) and strangeness quantum numbers ($S$) in an ideal gas in SB limit.}  \label{tab:2}
\end{center}
\end{table}  

In the next section, we focus on the second-order fluctuations and correlations of the conserved charges.

\subsubsection{Second-Order Moments} 
\label{2ndfluctuations}

\begin{figure}[htb]
\centering{
\includegraphics[width= 15 cm, height=10 cm, angle=0 ]{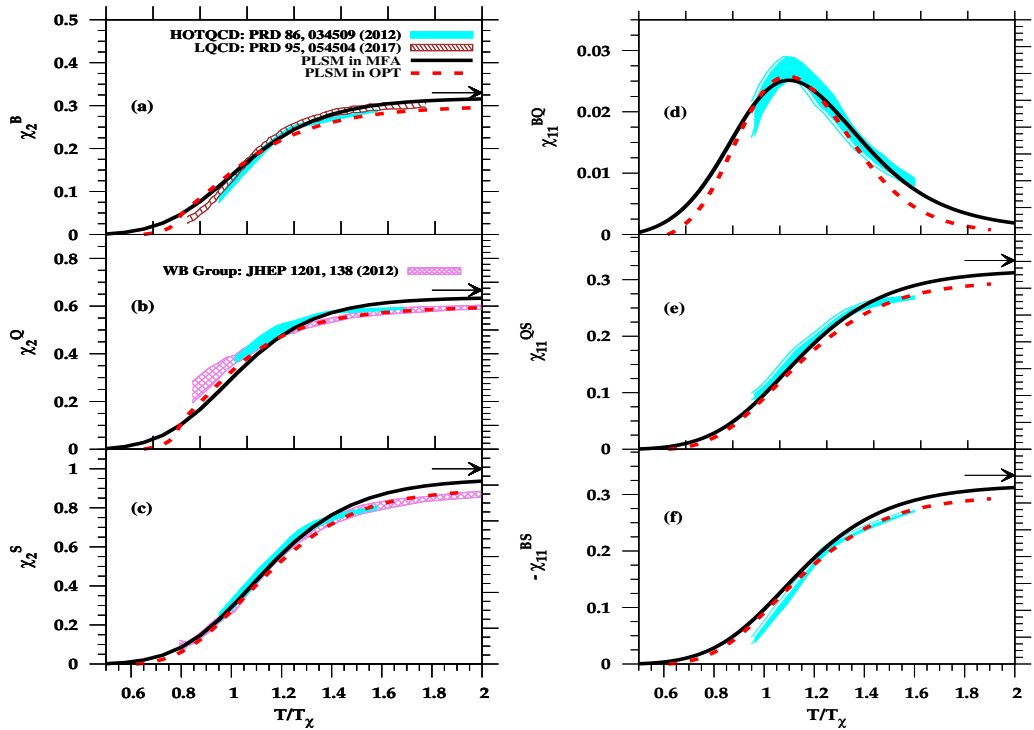}
\caption{\footnotesize (Color online) Left-hand panels illustrates the temperature dependence of the second-order fluctuations of conserved charges $\chi_2^B$, $\chi_2^Q$, and $\chi_2^S$ in panels (a), (b), and (c), respectively. The PLSM results in MFA (solid curve) and in OPT (dashed curve) are confronted to recent LQCD calculations \cite{Borsanyi:2011sw} (grid bands) \cite{Bazavov:2012jq} (solid bands) and \cite{Bazavov:2017dus} (dash band). Right-hand panels show the same as in the left-hand panel but here for the correlations of conserved charges $\chi_{11}^{BQ}$, $\chi_{11}^{QS}$ and $\chi_{11}^{BS}$ in panel (d), (e), (f), respectively.  
\label{Fig.2nd_flac_corrl}}
}
\end{figure}

With this regard, the PLSM results for the second-order fluctuations are deduced at $i+j+k=2$. In the calculations presented in Fig. \ref{Fig.2nd_flac_corrl}, we consider the MFA and OPT approaches at $\mu_f=0$. The PLSM results are compared with recent LQCD calculations. 

Figure \ref{Fig.2nd_flac_corrl} depicts the temperature dependence of the quadratic fluctuations $\chi_2^X$ (left-hand panel) and the quadratic correlations  $\chi_{11}^{XY}$  (right-hand panel) as calculated in the PLSM in MFA (solid curves) and in OPT (dashed curves) as functions of $T$ at vanishing chemical potential. Our calculations, the diagonal fluctuations, $\chi_{2}^{X}$, are compared with recent LQCD calculations; Wuppertal-Budapest (grid-filled bands) \cite{Borsanyi:2011sw} and  HotQCD \cite{Bazavov:2012jq} (solid-filled bands) and \cite{Bazavov:2017dus} (solid curve). The left panels of Fig. \ref{Fig.2nd_flac_corrl} show the susceptibilities for the net-baryon numbers, the net-electric charge, and the net-strangeness in panels (a), (b), and (c), respectively. The phase transition (crossover) is defined where the mesonic degrees of freedom start to be liberated from the confined phase (bound mesons) and the system is converted to the deconfined (free quarks and gluons) phase \cite{Borsanyi:2011bm, Cheng:2008zh}

It is obvious that the susceptibilities seem to have small values, at low temperatures, where in this region the mesonic contributions become dominant. Accordingly, the chiral condensates have maximum values due to the relevant degrees of freedom responsible for the small fluctuations, Fig. \ref{Fig.OrderPrameters}. It is obvious that in the region of crossover the fluctuations rapidly arise with the increase in temperature. It should be noticed that the chiral structure of the mesonic states plays an essential role in the temperature dependence of the fluctuations and the quark number multiplicities. It is believed that the new state-of-matter, i.e. massless quark and gluons, is produced. 

At high temperatures, the system reaches the deconfined state and therefore the fluctuations raise from fixed values to maxima, even if slightly below the SB limit. The explicit calculations from PLSM - as well - reach $\sim 94.3\%$ and $\sim 88.8\%$ of their respective ideal gas limits in MFA and in OPT, respectively. We notice that the differences between the PLSM results and the corresponding SB limits vary according to different scenarios, for instance, several mechanisms of the inclusion dynamics and degrees of freedom of quark flavors and gluons. Furthermore, the differentiation with respect to $\mu_f$ and the various types of statistical errors in the numerical simulations contribute to the potential differences.

The right-hand panel of Fig. \ref{Fig.2nd_flac_corrl} illustrates the same as in the left-hand panel  but here for the off-diagonal fluctuations (correlations) of the conserved quantum charges  $\chi_{11}^{XY}$; $\chi_{11}^{BQ}$ (top), $\chi_{11}^{QS}$ (middle), and $-\chi_{11}^{BS}$ (bottom panel). We notice that the PLSM correlations agree well with the LQCD calculations with continuum extrapolation \cite{Bazavov:2012jq}. The variation in the temperature behavior of the correlations of the baryon number with the electric charge, $\chi_{11}^{BQ}$, panel (d), is dominated by the mesonic contributions, at low temperatures. $\chi_{11}^{BQ}$  arise almost exponentially with the increase in temperature. Again $\chi_{11}^{BQ}$ vanishes, at high temperatures, because the quarks become massless in this limit \cite{Borsanyi:2011bm, Cheng:2008zh}. While the correlations of the strangeness $S$ with the electric charge $Q$ and that of the strangeness $S$  with the baryon number $B$;  $\chi_{11}^{QS}$ and  $-\chi_{11}^{BS}$, respectively, express clearly the effect of the strangeness degrees of freedom \cite{Koch:2005vg, Jeon:2000wg}. These correlations  are related to the quark-flavor fluctuations as $\chi_{11}^{BS}=-(\chi_2^s +2 \chi_{11}^{us})/3$ and $\chi_{11}^{QS}=(\chi_2^s - \chi_{11}^{us})/3$ \cite{Bazavov:2012jq}. As emphasized when discussing on the quadratic fluctuations, the correlations here are also sensitive to the quark-hadron contributions. We observe that the PLSM results agree well with the continuum extrapolation of the LQCD calculations, especially at low temperature and withing the region of crossover. At high temperatures, the PLSM results on $\chi_{11}^{QS}$ and  $-\chi_{11}^{BS}$ are below the SB limit. The difference is about $6\%$ and $12.2\%$ for MFA and OPT, respectively. 

We conclude that the comparison between the PLSM results and the LQCD calculations shows that OPT reproduces the first-principle fluctuations better than MFA, especially within the region of the phase transition. Also, OPT reproduces well the available LQCD correlations better that MFA.

\subsubsection{Higher-Order Moments} 
\label{Higherdfluctuations}

\begin{figure}[htb]
\includegraphics[width=16cm, height=4.5cm, angle=0]{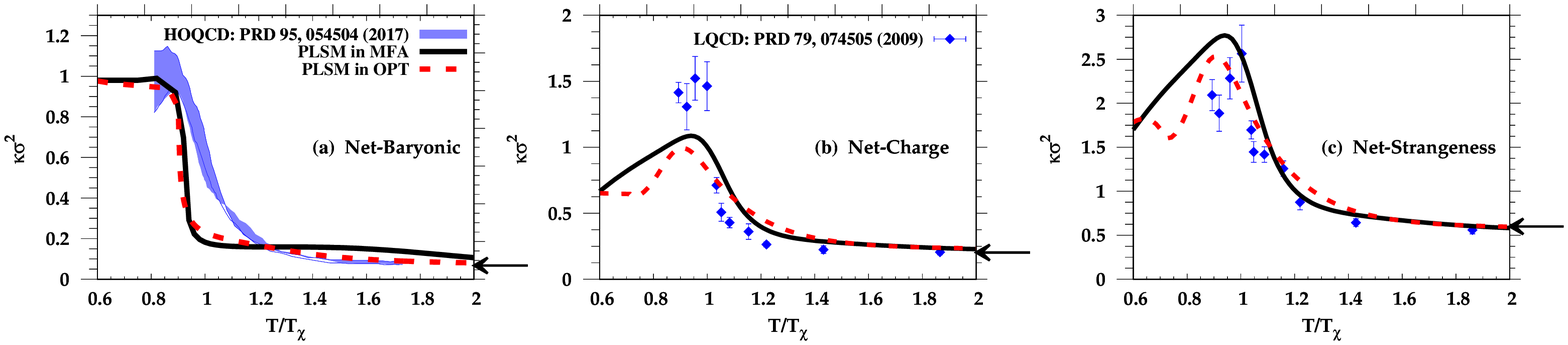}
\caption{\footnotesize (Color online) The temperature dependence of the moment products $\kappa\; \sigma^2$ computed in PLSM in MFA (solid curve) and in OPT (dashed curve) for net-baryon (a), net-charge (b) and net-strangeness (c) multiplicities and compared with the LQCD calculations by HotQCD collaborations (solid band)  \cite{Bazavov:2017dus} and \cite{Cheng:2008zh} (closed symbols). The arrows refers to the respective SB limits. 
\label{Fig.kappsigm2}}
 \end{figure}

The ratios of the higher-order moments, such as $\chi_4^B/\chi_2^B$, $\chi_4^Q/\chi_2^Q$, $\chi_4^S/\chi_2^S$, $\chi_6^B/\chi_2^B$ and $\chi_8^B$, are proposed as experimental signatures for the phase transition \cite{Friman:2011pf}. The PLSM results on the higher-order moments of different quantum charges in MFA are compared with the ones in OPT. We also compare these with the LQCD simulations and available experimental results. We start first with the PLSM results on the product of the higher-order moments $\kappa\,\sigma^2$ or $\chi_4^X/\chi_2^X$, Eq. (\ref{kappsigma2}), for which measurements  at various beam energies, for instance ref. \cite{Adamczyk:2017wsl}, and LQCD calculations at finite temperature \cite{Bazavov:2017dus,Cheng:2008zh} are available, Fiq. \ref{Fig.kappsigm2}. 

The susceptibility of the net-baryon number are proposed as a signature for deconfining hadrons into colored massless quarks and gluons \cite{Stephanov:2008qz}. It was suggested that the particle number correlations are directly coupled to the variance of the order parameter and thus becomes sensitive to the correlation length $\xi$ in a universal manner that $\chi_n\sim \xi^{n(5-\eta)/2-3}$, where $\eta$ is a critical exponent to be defined from the universality class.

In Fig. \ref{Fig.kappsigm2}, $\kappa\,\sigma^2$ for net-baryon number (left panel), for electric charge (middle panel) and for strangeness (right panel) are depicted as functions of $T$, at $\mu_B=0~$MeV. The results are compared with the LQCD calculations \cite{Cheng:2008zh} (closed symbols) and HotQCD collaboration \cite{Bazavov:2017dus}. We find that at low temperatures $\chi_4^B/\chi_2^B$ starts from unity, where the mesonic contributions are dominant within the hadron phase. With the increase in $T$, a rapid drop in $\chi_4^B/\chi_2^B=\kappa\, \sigma^2$ takes place, due to the rapid increase in $\chi^B_2$ within the region of crossover, left panel of Fig. \ref{Fig.kappsigm2}.  At high temperatures, the temperature dependence of $\chi_4^B/\chi_2^B$ reaches a saturated plateau, even with much lower values than that in the confined phase. We notice that the PLSM results in MFA and OPT seem to nearly approach the SB limit, $\left.\chi_4^B/\chi_2^B\right|_{SB}=2/(3\pi^2)$. 

Middle and right-hand panels of  Fig. \ref{Fig.kappsigm2} draw the moment product $\kappa\sigma^2$ for net-electric charge and net-strangeness multiplicity, respectively, as calculated from the PLSM and compared with LQCD calculations (closed symbols) \cite{Cheng:2008zh}. It seems that the $\kappa \sigma^2$ approaches unity at low temperatures and also the SB limit at high temperatures, while at $T \sim T_\chi$, there is an obvious peak observed in both quantum charges \cite{Cheng:2008zh}. At higher temperatures, $\kappa \sigma^2$ for net-charge agrees with the LQCD simulations, while that of net-strangeness multiplicity seems to have a different dependence, at least within the region of temperatures covered by the lattice calculations. The lattice calculations we compare with is about 10 years old. To the authors' best knowledge, so far no recent calculations are available to compare with. There is a qualitative agreement, especially at high temperatures so that the results approach the SB limits $2/\pi^2$ and $6/\pi^2$ for net-electric charge and net-strangeness multiplicities, respectively.

\begin{figure}[htb]
\centering{
\includegraphics[width= 16 cm, height=7 cm, angle=0 ]{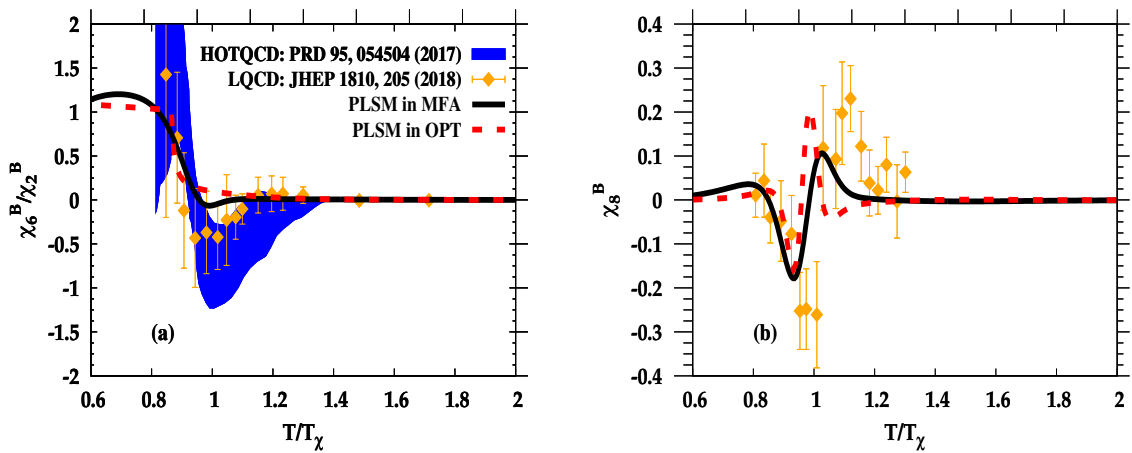}
\caption{\footnotesize (Color online) The temperature dependence of  susceptibilities of conserved charges $\chi_6^B/\chi_2^B$ and $\chi_8^B$ computed in PLSM in MFA (solid curve) and in OPT (dashed curve). The LQCD calculations \cite{Borsanyi:2018grb} (solid symbols) and  \cite{Bazavov:2017dus} (bands). 
\label{Fig.higermoments}}
}
 \end{figure}

The comparison between the PLSM results in MFA and in OPT is qualitatively possible when confronting both results with the continuum extrapolation of LQCD results \cite{Bazavov:2017dus}. In Fig. \ref{Fig.higermoments}, the temperature dependence of susceptibilities of conserved charges $\chi_6^B/\chi_2^B$ (left panel) and $\chi_8^B$ (right panel) computed in PLSM in MFA (solid curve) and in OPT (dashed curve) is compared with the LQCD calculations \cite{Borsanyi:2018grb} (solid symbols) and  \cite{Bazavov:2017dus} (bands). 

For $\chi_6^B/\chi_2^B$ (left panel), at $T/T_{\chi}\gtrsim 1.2$, although the vanishing temperature dependence, the PLSM results agree excellently with the LQCD calculations. At lower temperatures, we find a minimum at $T_{\chi}$ followed by a rapid arise. This continues to a maximum at $\chi_4^B/\chi_2^B\sim 1$. The LQCD calculations are limited to $0.8 T_{\chi}$. The temperature dependence of these quantities are significant for the quark-hadron phase transition \cite{Borsanyi:2011bm, Cheng:2008zh}. Within their large errorbars, we conclude that both PLSM approximations, MFA and OPT, at least qualitatively, reproduce well the LQCD calculations.  

For $\chi_8^B$ (right panel), the excellent agreement at $T/T_{\chi}\gtrsim 1.2$ exists here, as well. Decreasing the temperature unveil an interesting structure. A sinusoidal dependence is obtained, the sinusoidal oscillation is also supported by the LQCD calculations \cite{Borsanyi:2018grb}. The most remarkable features of such PLSM results in both MFA and OPT indicate a smooth crossover between the hadronic and partonic phases. At $T/T_{\chi}\lesssim 0.8$,  $\chi_8^B$ vanishes. The structure produced by OPT fits well with the LQCD. 

So far, we conclude that OPT is more sensitive to the higher-order moment products $\chi_6^B/\chi_2^B$, and $\chi_8^B$ than the MFA. The off-diagonal cumulants of net-proton, net-charge, and net-kaon multiplicity distributions are measured in the STAR experiment at energies ranging between $7.7$ and $200~$GeV within the rapidity $|y|<0.5$ and the transverse momentum range $0.4 <p_T<2.0~$ GeV \cite{Esha:2017dce,Esha:2017yeh,Chen:2016xyu}.

\section{Conclusions \label{conclusion}} 

We have generalized PLSM to incorporate zero- and higher-order $\delta$-expansions, i.e. MFA and OPT. While MFA is originated in statistical physics and  can be applied to a wide spectrum of numerical implications, OPT was developed in $\mathcal{O}(N)\; \phi^4$ theory in order to resum the higher-order terms of the naive perturbation approach. We aim at appointing whether the MFA-generalization reproduces identical results or eventually improves the PLSM calculations in OPT. When improvements are observed, we aim at highlighting the high-order moments affected by OPT.

At finite temperatures and chemical potentials, we have compared the chiral condensates and the decofinement order parameters, the thermodynamic pressure, the subtracted condensates, the pseudo-critical temperatures, the second- and high-order moments of various conserved charges (cumulants) obtained in MFA with the ones obtained in OPT. These calculations are then confronted to available lattice QCD simulations. In general, we conclude that when moving to lower- to higher-order moments, the OPT approach becomes more and more reliable than the MFA. For example, we found that OPT is more sensitive to $\chi_6^B/\chi_2^B$, and $\chi_8^B$ than the MFA.

The convincing OPT results outlined in this study encourages the trial to compare with the measurements that are and shall be available in the near future. Accordingly, some PLSM parameters can be adjusted, properly, on one hand. On the other hand, the critical phenomena that might be detected in the experimental results and are sensitive the higher-order moments could likely be predicted.

\section*{Acknowledgements}
The work of AT was supported by the ExtreMe Matter Institute EMMI at the GSI Helmholtz Centre for Heavy Ion Research.

%
\bibliographystyle{aip}
\bibliography{PLSM_OPTMFA_V2} 

\begin{thebibliography}{10}

\bibitem{Schwinger:1951xk}
J.~S. Schwinger,
\newblock Phys. Rev. {\bf 82}, 914 (1951).

\bibitem{Schwinger:1953tb}
J.~S. Schwinger,
\newblock Phys. Rev. {\bf 91}, 713 (1953).

\bibitem{Schwinger:1953zza}
J.~Schwinger,
\newblock Phys. Rev. {\bf 91}, 728 (1953).

\bibitem{Schwinger:1953zz}
J.~Schwinger,
\newblock Phys. Rev. {\bf 92}, 1283 (1953).

\bibitem{Schwinger:1954zza}
J.~Schwinger,
\newblock Phys. Rev. {\bf 93}, 615 (1954).

\bibitem{Schwinger:1954zz}
J.~Schwinger,
\newblock Phys. Rev. {\bf 94}, 1362 (1954).

\bibitem{GellMann:1960np}
M.~Gell-Mann and M.~Levy,
\newblock Nuovo Cim. {\bf 16}, 705 (1960).

\bibitem{Schwinger:1957em}
J.~S. Schwinger,
\newblock Annals Phys. {\bf 2}, 407 (1957).

\bibitem{Birse:1994cz}
M.~C. Birse,
\newblock J. Phys. {\bf G20}, 1537 (1994).

\bibitem{Roder:2003uz}
D.~Roder, J.~Ruppert, and D.~H. Rischke,
\newblock Phys. Rev. {\bf D68}, 016003 (2003).

\bibitem{Gallas:2009qp}
S.~Gallas, F.~Giacosa, and D.~H. Rischke,
\newblock Phys. Rev. {\bf D82}, 014004 (2010).

\bibitem{Tawfik:2014gga}
A.~N. Tawfik and A.~M. Diab,
\newblock Phys. Rev. {\bf C91}, 015204 (2015).

\bibitem{Wesp:2017tze}
C.~Wesp, H.~van Hees, A.~Meistrenko, and C.~Greiner,
\newblock Eur. Phys. J. {\bf A54}, 24 (2018).

\bibitem{Tawfik:2016gye}
A.~N. Tawfik, A.~M. Diab, and M.~T. Hussein,
\newblock J. Phys. {\bf G45}, 055008 (2018).

\bibitem{AbdelAalDiab:2018hrx}
A.~M. Abdel Aal~Diab and A.~N. Tawfik,
\newblock EPJ Web Conf. {\bf 177}, 09005 (2018).

\bibitem{Tawfik:2019rdd}
A.~N. Tawfik, A.~M. Diab, and M.~T. Hussein,
\newblock Chin. Phys. {\bf C43}, 034103 (2019).

\bibitem{Tawfik:2016lih}
A.~N. Tawfik, A.~M. Diab, N.~Ezzelarab, and A.~G. Shalaby,
\newblock Adv. High Energy Phys. {\bf 2016}, 1381479 (2016).

\bibitem{Tawfik:2016ihn}
A.~N. Tawfik, A.~M. Diab, and T.~M. Hussein,
\newblock Int. J. Adv. Res. Phys. Sci. {\bf 3}, 4 (2016).

\bibitem{Tawfik:2017cdx}
A.~N. Tawfik, A.~M. Diab, and M.~T. Hussein,
\newblock J. Exp. Theor. Phys. {\bf 126}, 620 (2018).

\bibitem{Tawfik:2016edq}
A.~N. Tawfik, A.~M. Diab, and M.~T. Hussein,
\newblock Int. J. Mod. Phys. {\bf A31}, 1650175 (2016).

\bibitem{Tawfik:2019tkp}
A.~N. Tawfik, A.~M. Diab, M.~T. Ghoneim, and H.~Anwer,
\newblock (2019).

\bibitem{Randrup:1996es}
J.~Randrup,
\newblock Nucl. Phys. {\bf A616}, 531 (1997).

\bibitem{Kunihiro:1983ej}
T.~Kunihiro and T.~Hatsuda,
\newblock Prog. Theor. Phys. {\bf 71}, 1332 (1984).

\bibitem{Thoma:1989ip}
M.~H. Thoma and H.~J. Mang,
\newblock Z. Phys. {\bf C44}, 349 (1989).

\bibitem{Thoma:1989in}
M.~H. Thoma,
\newblock Z. Phys. {\bf C44}, 343 (1989).

\bibitem{Weinberg:1995mt}
S.~Weinberg,
\newblock {\em {The Quantum theory of fields. Vol. 1: Foundations}},
\newblock Cambridge University Press, 2005.

\bibitem{Schaefer:2008hk}
B.-J. Schaefer and M.~Wagner,
\newblock Phys. Rev. {\bf D79}, 014018 (2009).

\bibitem{Stevenson:1981vj}
P.~M. Stevenson,
\newblock Phys. Rev. {\bf D23}, 2916 (1981).

\bibitem{Stevenson:1981rz}
P.~M. Stevenson,
\newblock Phys. Rev. {\bf D24}, 1622 (1981).

\bibitem{Kneur:2010yv}
J.-L. Kneur, M.~B. Pinto, and R.~O. Ramos,
\newblock Phys. Rev. {\bf C81}, 065205 (2010).

\bibitem{Schaefer:2009ui}
B.-J. Schaefer, M.~Wagner, and J.~Wambach,
\newblock Phys. Rev. {\bf D81}, 074013 (2010).

\bibitem{Mao:2009aq}
H.~Mao, J.~Jin, and M.~Huang,
\newblock J. Phys. {\bf G37}, 035001 (2010).

\bibitem{Klevansky:1992qe}
S.~P. Klevansky,
\newblock Rev. Mod. Phys. {\bf 64}, 649 (1992).

\bibitem{Dmitrasinovic:1995cb}
V.~Dmitrasinovic, H.~J. Schulze, R.~Tegen, and R.~H. Lemmer,
\newblock Annals Phys. {\bf 238}, 332 (1995).

\bibitem{Oertel:1999fk}
M.~Oertel, M.~Buballa, and J.~Wambach,
\newblock Phys. Lett. {\bf B477}, 77 (2000).

\bibitem{Restrepo:2014fna}
T.~E. Restrepo, J.~C. Macias, M.~B. Pinto, and G.~N. Ferrari,
\newblock Phys. Rev. {\bf D91}, 065017 (2015).

\bibitem{Hansen:2006ee}
H.~Hansen et~al.,
\newblock Phys. Rev. {\bf D75}, 065004 (2007).

\bibitem{Costa:2008dp}
P.~Costa, M.~C. Ruivo, C.~A. de~Sousa, H.~Hansen, and W.~M. Alberico,
\newblock Phys. Rev. {\bf D79}, 116003 (2009).

\bibitem{Cea:2003un}
P.~Cea, L.~Cosmai, and M.~D'Elia,
\newblock Nucl. Phys. Proc. Suppl. {\bf 129}, 751 (2004),
\newblock [,751(2003)].

\bibitem{Haas:2013qwp}
L.~M. Haas, R.~Stiele, J.~Braun, J.~M. Pawlowski, and J.~Schaffner-Bielich,
\newblock Phys. Rev. {\bf D87}, 076004 (2013).

\bibitem{GellMann:1968rz}
M.~Gell-Mann, R.~J. Oakes, and B.~Renner,
\newblock Phys. Rev. {\bf 175}, 2195 (1968).

\bibitem{Dashen:1969eg}
R.~F. Dashen,
\newblock Phys. Rev. {\bf 183}, 1245 (1969).

\bibitem{Borsanyi:2010bp}
S.~Borsanyi et~al.,
\newblock JHEP {\bf 09}, 073 (2010).

\bibitem{Bellwied:2015rza}
R.~Bellwied et~al.,
\newblock Phys. Lett. {\bf B751}, 559 (2015).

\bibitem{Bazavov:2018mes}
A.~Bazavov et~al.,
\newblock Phys. Lett. {\bf B795}, 15 (2019).

\bibitem{Das:2014qca}
S.~Das,
\newblock (2014),
\newblock [EPJ Web Conf.90,08007(2015)].

\bibitem{Andronic:2017pug}
A.~Andronic, P.~Braun-Munzinger, K.~Redlich, and J.~Stachel,
\newblock Nature {\bf 561}, 321 (2018).

\bibitem{Borsanyi:2016ksw}
S.~Borsanyi et~al.,
\newblock Nature {\bf 539}, 69 (2016).

\bibitem{Kapusta:2006book}
J.~I. Kapusta and C.~Gale,
\newblock {\em {Finite-temperature field theory: Principles and applications}},
\newblock Cambridge University Press, 2006,
\newblock UK.

\bibitem{Luo:2017faz}
X.~Luo and N.~Xu,
\newblock Nucl. Sci. Tech. {\bf 28}, 112 (2017).

\bibitem{Borsanyi:2011sw}
S.~Borsanyi et~al.,
\newblock JHEP {\bf 01}, 138 (2012).

\bibitem{Bazavov:2012jq}
A.~Bazavov et~al.,
\newblock Phys. Rev. {\bf D86}, 034509 (2012).

\bibitem{Bazavov:2017dus}
A.~Bazavov et~al.,
\newblock Phys. Rev. {\bf D95}, 054504 (2017).

\bibitem{Borsanyi:2011bm}
S.~Borsanyi et~al.,
\newblock J. Phys. {\bf G38}, 124060 (2011).

\bibitem{Cheng:2008zh}
M.~Cheng et~al.,
\newblock Phys. Rev. {\bf D79}, 074505 (2009).

\bibitem{Koch:2005vg}
V.~Koch, A.~Majumder, and J.~Randrup,
\newblock Phys. Rev. Lett. {\bf 95}, 182301 (2005).

\bibitem{Jeon:2000wg}
S.~Jeon and V.~Koch,
\newblock Phys. Rev. Lett. {\bf 85}, 2076 (2000).

\bibitem{Friman:2011pf}
B.~Friman, F.~Karsch, K.~Redlich, and V.~Skokov,
\newblock Eur. Phys. J. {\bf C71}, 1694 (2011).

\bibitem{Adamczyk:2017wsl}
L.~Adamczyk et~al.,
\newblock Phys. Lett. {\bf B785}, 551 (2018).

\bibitem{Stephanov:2008qz}
M.~A. Stephanov,
\newblock Phys. Rev. Lett. {\bf 102}, 032301 (2009).

\bibitem{Borsanyi:2018grb}
S.~Borsanyi et~al.,
\newblock JHEP {\bf 10}, 205 (2018).

\bibitem{Esha:2017dce}
R.~Esha,
\newblock Nucl. Phys. {\bf A967}, 457 (2017).

\bibitem{Esha:2017yeh}
R.~Esha,
\newblock PoS {\bf CPOD2017}, 003 (2018).

\bibitem{Chen:2016xyu}
L.~Chen, Z.~Li, F.~Cui, and Y.~Wu,
\newblock Nucl. Phys. {\bf A957}, 60 (2017).

\end{thebibliography}

\end{document}